\documentclass[prb,twocolumn,groupedaddress,floatfix]{revtex4-2}
\usepackage{amsmath}

\usepackage{lipsum} 
\usepackage{verbatim}
\usepackage{wasysym} 
\usepackage{epsfig}
\usepackage{subfigure}
\usepackage{graphicx}
\usepackage{amsfonts}
\usepackage[figuresright]{rotating}
\usepackage{amssymb}
\usepackage{amsmath}
\usepackage{psfrag}
\usepackage{esint} 
\usepackage{bm}
\usepackage[colorlinks,linkcolor=blue,anchorcolor=blue,citecolor=blue,urlcolor=blue]{hyperref}
\usepackage{graphicx}
\usepackage[version=4]{mhchem}

\usepackage{verbatim}

\def\be{\begin{equation}} \def\ee{\end{equation}}
\def\bea{\begin{eqnarray}} \def\eea{\end{eqnarray}}

\def\k{{\bf k}}

\renewcommand{\vec}[1]{\mathbf{#1}}

\def\bpm{\begin{pmatrix}} \def\epm{\end{pmatrix}}

\renewcommand{\vr}{\mathbf{r}}

\makeatletter
\newcommand*{\balancecolsandclearpage}{%
  \close@column@grid
  \clearpage
}
\makeatother

\begin{document}

\title{Robust Flat Bands with Tunable Energies in Honeycomb Superlattices}

\author{Zihao Qi}
\affiliation{Department of Physics, Math and Astronomy, California Institute of Technology, Pasadena, California 91125, USA}
\author{Eric Bobrow}
\affiliation{Department of Physics and Astronomy, Johns Hopkins University, Baltimore, Maryland 21218, USA}
\author{Yi Li}
\email{yili.phys@jhu.edu}
\affiliation{Department of Physics and Astronomy, Johns Hopkins University, Baltimore, Maryland 21218, USA}

\date{December 14, 2020}

\begin{abstract}
Flat bands in lattice models have provided useful platforms for studying strong correlation and topological physics. 
Recently, honeycomb superlattices have been shown to host flat bands that persist in the presence of local perturbations respecting lattice symmetries. We analytically derive the flat band energies in the presence of longer range hopping and find that the energies of flat bands are \textit{tunable} by these perturbations. In real space, the wave function is constructed from standing waves on each honeycomb edge, allowing the construction of plaquette and loop eigenstates due to destructive interference in real space that give rise to the flat bands robust against long range hoppings. 
\end{abstract}

\maketitle

\section{Introduction}
\label{sec:intro}
Recently, there has been intense interest in flat band physics. 
When electrons fill a completely dispersionless band, interaction effects become non-perturbative, and novel strongly-correlated phases and phenomena can be developed \cite{Goda2006, Wu2007, Bergman2008, Bistritzer2011, Tang2011, Wang2013, Po2018,  Zou2018, Liu2019a, Tarnopolsky2019, Volovik2019, Chiu2020}. 
In realistic materials, due to unavoidable further-than-nearest-neighbor hoppings, flat bands often develop dispersion. 
It would be valuable to search for flat bands that are robust in the presence of perturbations from longer range hoppings. 

Motivated by a recent scanning tunnelling microscopy experiment on the nearly commensurate charge-density wave phase of 1T-TaS$_2$ \cite{GYCold}, where metallic states are formed along imperfect David-star clusters of the domain wall network that form a honeycomb superlattice, flat bands have been discovered in the corresponding tight-binding models on the superlattice. The large density of states from flat bands can contribute to enhanced superconductivity and give rise to other correlation effects. 
Remarkably, the tight-binding models defined on honeycomb superlattices with each honeycomb edge containing multiple sites turn out to host flat bands robust against symmetric perturbations from longer range hoppings \cite{GYC}. 

Sharing the same physics as the familiar examples of flat bands in Lieb lattice \cite{Xia2018},
line graphs \cite{Mielke1991} such as the kagom\'e lattice, or graphs constructed from complete subgraphs \cite{Tanaka2020}, the honeycomb superlattice flat bands arise from macroscopically degenerate states localized 
around plaquettes and loops of the lattice as a consequence of 
destructive interference of hopping amplitudes. 
The honeycomb superlattice differs from the line graphs as well as the Lieb lattice in the structure of the complete subgraphs to which sites on two edges of a plaquette supporting localized states are connected.
We find that the honeycomb superlattice demonstrates an interesting 
example of a general family of lattices supporting flat bands 
when the sites from two neighboring edges are connected to multiple 
complete subgraphs as long as these subgraphs are disconnected, 
inter-connected by additional bonds, or with a single site between 
two additional bonds outside of the plaquette with localized states. 
Our results could help understand and predict flat bands in realistic materials in the presence of longer range hoppings. 

The remainder of the paper is organized as follows: in Sec. \ref{sec:model}, we introduce the hopping model on the honeycomb superlattice. 
In Sec. \ref{sec:fb}, we analytically derive the flat band energies and show their tunability. 
In Sec. \ref{sec:localizedstates}, we analytically derive and construct the standing waves and localized states living on hexagonal plaquettes. 
Finally, we summarize our paper in Sec. \ref{sec:conclusion}.

\section{\label{sec:model} Flat bands on a honeycomb superlattice with nearest neighbor hopping}
We start by reviewing the tight-binding model used by Cho and his collaborators [\onlinecite{GYC}], who discovered that a family of honeycomb superlattices can support flat bands in the presence of longer-range hoppings and spin-orbit couplings. 

In the presence of only nearest-neighbor (N.N.) hoppings, the spinless tight-binding Hamiltonian is
\begin{equation}
    H_{\mathrm{N.N.}} = t_0 \sum_{\langle \mathbf{r}, \mathbf{r'}\rangle} c_{\mathbf{r}}^{\dagger} c_{\mathbf{r}'} + h.c.,
\label{eq:Hamiltonian}
\end{equation}
where the sum runs over the pairs of N.N. sites $\langle \mathbf{r},\mathbf{r}' \rangle$,
$c_{\mathbf{r}}^{\dagger}$ ($c_{\mathbf{r}}$) is the creation (annihilation) operator of an electron at site $\mathbf{r}$, and $t_0$ is the amplitude of the hopping between N.N sites in the honeycomb superlattice where each honeycomb edge is decorated with additional sites. 
We take $t_0 = 1$ with positive sign, since the model with only N.N. hoppings is bipartite. We also take the spacing between two N.N. sites to be $1$ throughout. 

\begin{figure}[htpb]
\subfigure[]{\includegraphics[scale=0.2]{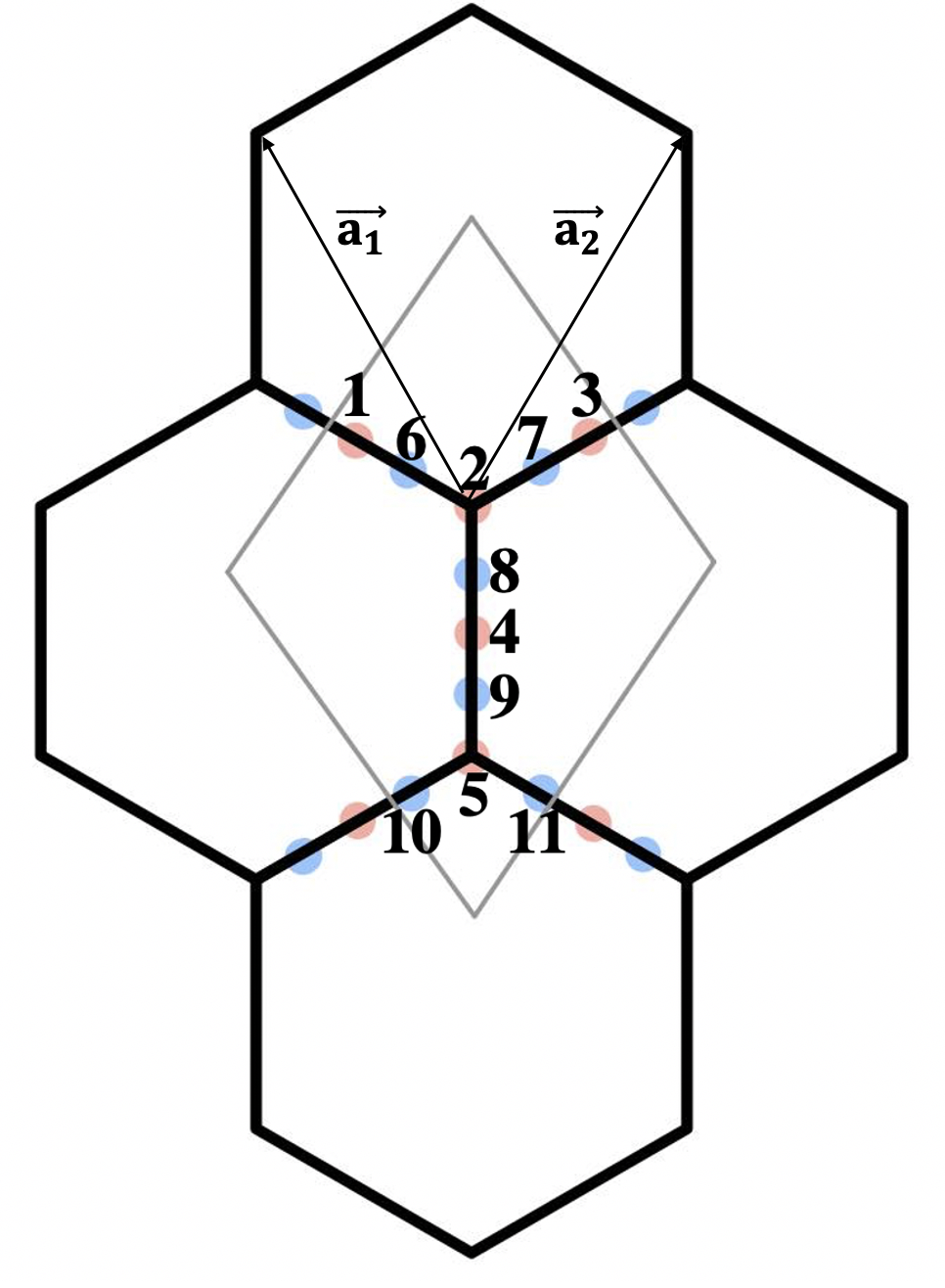}}
\subfigure[]{\includegraphics[scale=0.19]{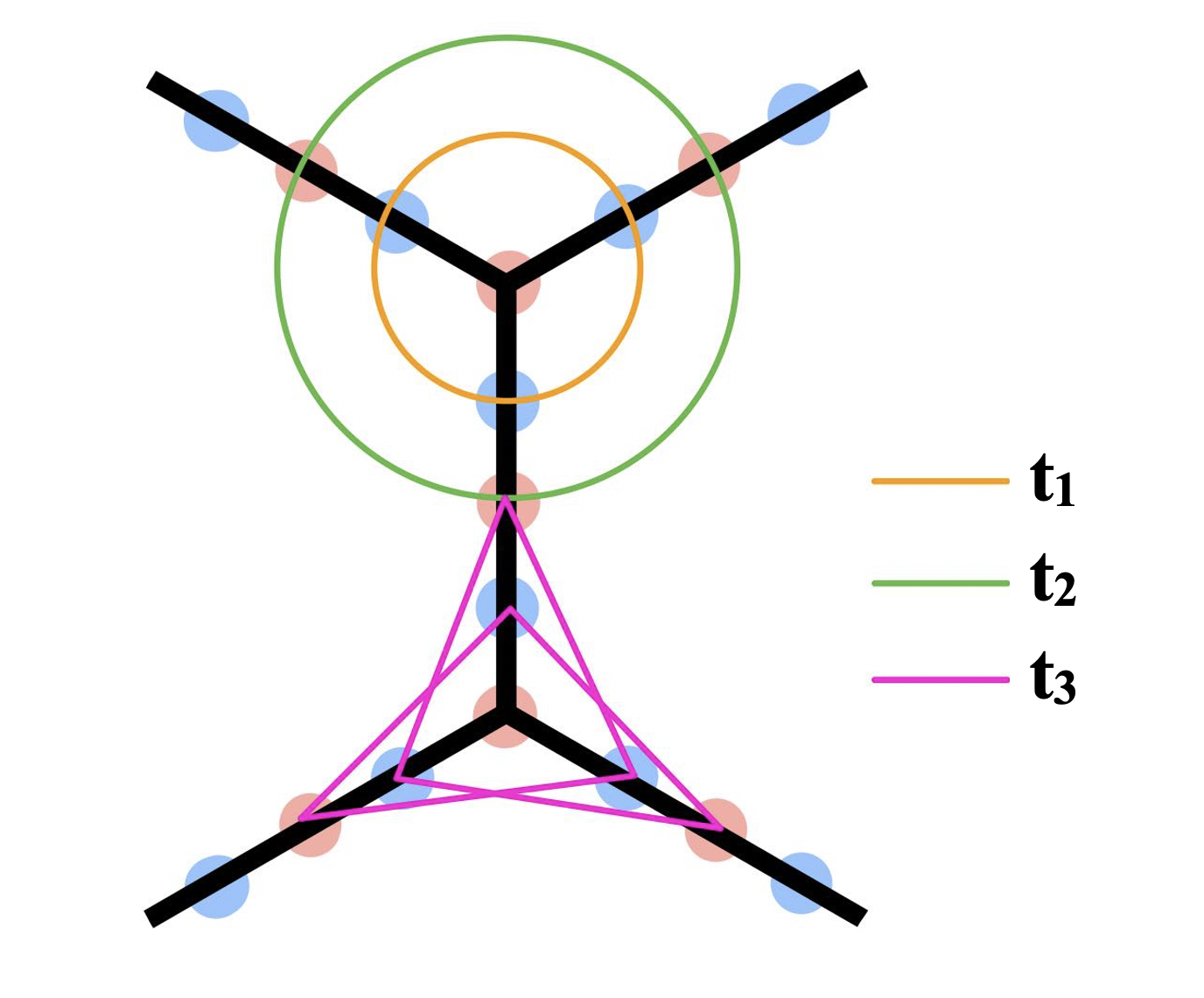}}
\caption{(a) The unit cell of a honeycomb superlattice. The sites in red, numbered from 1 to 5, are in the $A$ sublattice, whereas the sites in blue, numbered from 6 to 11, belong to the $B$ sublattice. The primitive lattice vectors $\vec{a_1}$ and $\vec{a_2}$ are also labelled.
(b) The set of $C_3$-symmetric perturbations we introduce. The circular hoppings in orange and green have amplitudes $t_1$ and $t_2$, whereas the triangular perturbations in purple have amplitude $t_3$.}
\label{fig:model}
\end{figure}

We consider the honeycomb superlattice with five sites per honeycomb edge, which is described by the eleven-site unit cell shown in Fig. \ref{fig:model}(a). 
Similar results can be obtained for superlattices with different numbers of additional sites.
Following the numbering scheme in Fig. \ref{fig:model}(a), the $A$ sublattice consists of two vertices and three bond-center sites numbered 1 through 5, while the rest of the sites, numbered 6 to 11, belong to the $B$ sublattice. 
The system exhibits three-fold rotation symmetry with respect to any vertex of the superlattice and mirror symmetries with mirror planes going through opposite vertices and opposite edge centers of a honeycomb plaquette.

In the momentum space representation, the Hamiltonian can be written
\begin{equation}
    H_{\mathrm{N.N.}}  =\sum_{\vec{k}} \sum_{i, j = 1}^{11} c^\dagger_{\vec{k},i} \left[\mathcal{H}_{\mathrm{N.N.}}(\k)\right]_{ij} c_{\vec{k},j}
\end{equation}
with $\mathbf{k} = (k_x, k_y)$ and
$c^\dagger_{\vec{k}, i} = 1/\sqrt{N_0} \sum_{\vec{R}} e^{i \vec{k}\cdot\vec{R}} c^\dagger_{\vec{R}, i}$, where $c^\dagger_{\vec{R}, i}$ is the real space creation operator at site $i$ of unit cell $\vec{R}$ and $\vec{R}$ is summed over all $N_0$ unit cells. 
Due to the bipartite hopping, the kernel   ${\mathcal{H}}_{\mathrm{N.N.}}(\vec{k})$ 
takes the block off-diagonal form
\begin{equation}
{\mathcal{H}}_{\mathrm{N.N.}}(\vec{k}) =  \begin{bmatrix}
0  & {\mathcal{H}}_{BA}(\vec{k}) \\
{\mathcal{H}}_{AB}(\vec{k})  & 0 \\
  \end{bmatrix}.
\label{eq:matrix}
\end{equation}
Here, $0$ denotes a zero rectangular matrix of appropriate dimensions, and  ${\mathcal{H}}_{AB}={\mathcal{H}}_{BA}^{\dagger}$ describes the N.N. hoppings from $A$ to $B$ sublattices with
\begin{equation}
    {\mathcal{H}}_{BA}(\vec{k}) = \begin{bmatrix}
  t_0 & 0 & 0 & 0 & 0 & t_0e^{-i\mathbf{k} \cdot \mathbf{a_1}} \\
  t_0 & t_0 & t_0 & 0 & 0 & 0 \\
  0 & t_0 & 0 & 0 & t_0e^{-i\mathbf{k} \cdot \mathbf{a_2}} & 0 \\
  0 & 0 & t_0 & t_0 & 0 & 0\\
  0 & 0 & 0 & t_0 & t_0 & t_0
  \end{bmatrix},
\label{eq:HBA}
\end{equation}
where $\mathbf{a_1} = (-2 \sqrt{3}, 6) $ and $\mathbf{a_2} = (2 \sqrt{3}, 6) $ are the primitive direct lattice vectors of a superlattice unit cell.
\begin{figure}[htpb]
\subfigure[]{\includegraphics[scale=0.245]{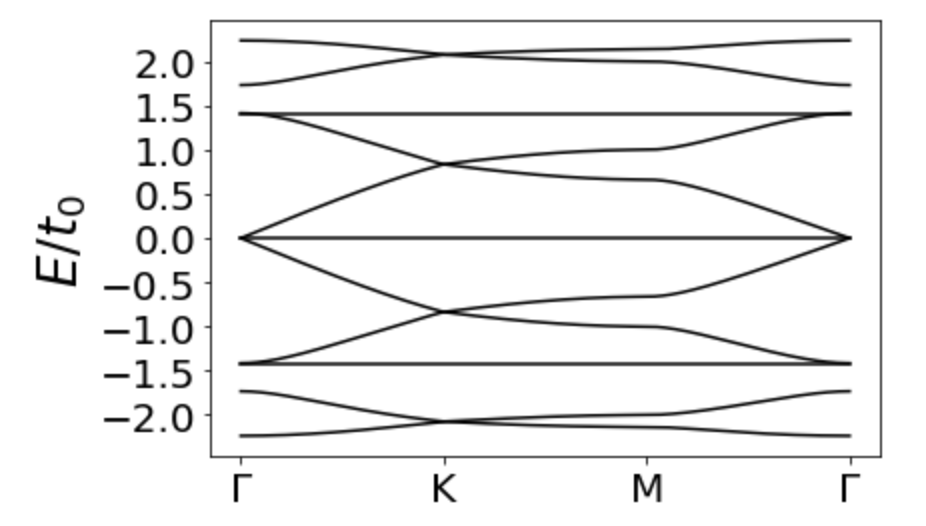}}
\subfigure[]{\includegraphics[scale=0.242]{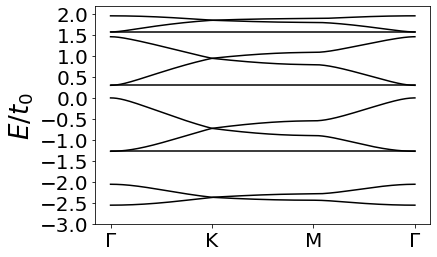}}
\subfigure[]{\includegraphics[scale=0.242]{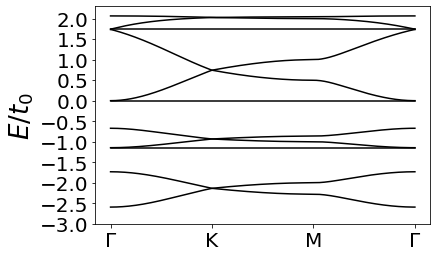}}
\subfigure[]{\includegraphics[scale=0.245]{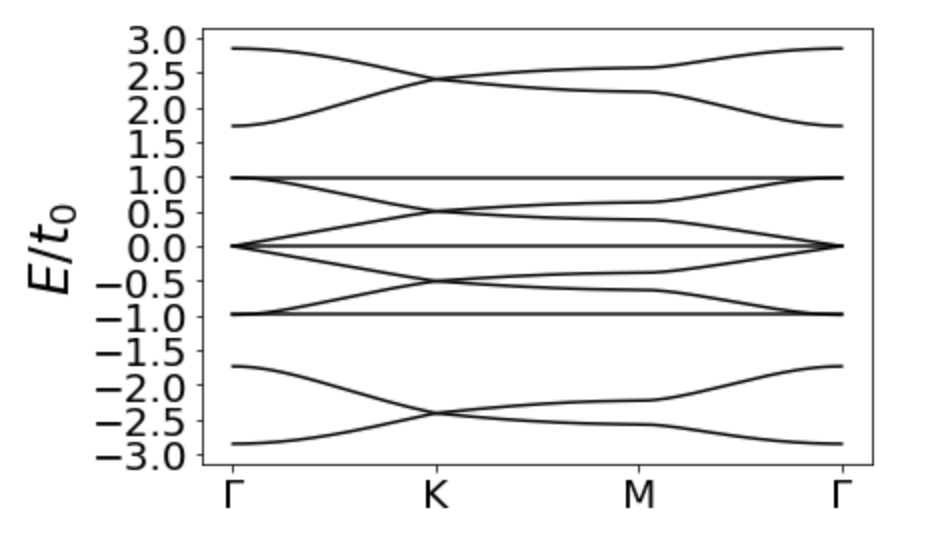}}
\caption{Band structures plotted along the high-symmetry cut $\Gamma$-$K$-$M$-$\Gamma$ for different sets of perturbations in units of $t_0$. The high-symmetry points in momentum space are $\Gamma = \mathbf{0}$, $K= \left( \frac{\pi}{6\sqrt{3}}, \frac{\pi}{6} \right)$, and $M= \left(0, \frac{\pi}{6}\right)$.  
The three flat bands remain robust for different parameters: (a) $t_1 = 0$,  $t_2 = 0$, $t_3 = 0$, \textit{i.e.}, only N.N. hoppings; (b) $t_1 = 0.3$,  $t_2 = 0$, $t_3 = 0$; (c) $t_1 = 0$, $t_2 = 0.3$, $t_3 = 0$; (d) $t_1 = 0$, $t_2 = 0$, $t_3 = 0.3$. Here (b) reproduces Fig. 1(c) in Ref. \onlinecite{GYC}.}
\label{fig:bandstructures}
\end{figure}



As shown in Fig. \ref{fig:bandstructures}(a), three out of the eleven bands of $H_{\mathrm{N.N.}}(\vec{k})$ described by Eq. (\ref{eq:matrix}) are \textit{completely flat} at energies $E_{\mathrm{flat, mid}} = 0$ and $E_{\mathrm{flat, top/bot}}=\pm \sqrt{2}$; we will derive these energies analytically in Section. \ref{sec:fb}. Note that the band structure is symmetric with respect to zero energy, indicating particle-hole symmetry, which follows from the transformation $\mathcal{C}$: $c_{\mathbf{r}} \rightarrow c_{\mathbf{r}}$ for $\vr \in$
\text{$A$ sublattice}; $c_{\mathbf{r}} \rightarrow - c_{\mathbf{r}}$ for $\vr \in$ \text{$B$ sublattice}. 

The bands in Fig. \ref{fig:bandstructures}(a) also exhibit crossings at K and $\Gamma$ points, most of which are Dirac type except the quadratic crossings at the $\Gamma$ point for the highest and the lowest flat bands.
Furthermore, because of the particle-hole symmetry, a three-fold degeneracy occurs at the $\Gamma$ point for the middle flat band.

Reference \onlinecite{GYC} found in numeric calculations that the flat band wave functions vanish on any honeycomb vertex. Here we provide an analytical  argument. 
In the basis of eleven sublattices, a general Bloch state $\psi(\vec{k}) = ( \psi_1(\vec{k}), \psi_2(\vec{k}), \cdots, \psi_{11}(\vec{k}))^{T}$ satisfies its eigenequation
\begin{equation}
    {\mathcal{H}}_{\mathrm{N.N.}}(\vec{k}) \psi(\vec{k}) = E \psi(\vec{k}).
\label{eq:schrodinger}
\end{equation}
Or, in terms of the eleven components of $\psi(\vec{k})$, we have
$\psi_6 + \psi_{11} e^{-i\vec{k} \cdot \vec{a_1}} = E \psi_1$, 
$\psi_6 + \psi_7 + \psi_8 = E \psi_2$,  
$\psi_7 + \psi_{10} e^{-i\vec{k} \cdot \vec{a_2}} = E \psi_3$, 
$\psi_8 + \psi_9 = E \psi_4$, 
$\psi_9 +\psi_{10} + \psi_{11} = E \psi_5$, 
$\psi_1 +\psi_2 = E\psi_6$, 
$\psi_2 + \psi_3 = E \psi_7$, 
$\psi_2 + \psi_4 = E \psi_8$, 
$\psi_4 +\psi_5 = E \psi_9 $, 
$e^{i\vec{k} \cdot \vec{a_2}} \psi_3+ \psi_5 = E \psi_{10}$, and 
$e^{i\vec{k} \cdot \vec{a_1}} \psi_1 + \psi_5 = E \psi_{11}$. 
Here, for brevity, we suppressed the momentum dependence of $\psi_i$. 

Note that the above set of equations manifestly satisfies the system's symmetries: it stays invariant under $C_3$ rotations about a honeycomb vertex and inversions about the mirror planes. Solving for $\psi_5(\vec{k})$ in terms of $\psi_2(\vec{k})$ yields $(E^4-5E^2+3) \psi_5(\vec{k}) =  (e^{-i\vec{k} \cdot \vec{a_1}} + e^{-i\vec{k} \cdot \vec{a_2}} + 1) \psi_2(\vec{k})$. Under the inversion about the horizontal plane going through site 4, $\psi_5(\vec{k}) \rightarrow \psi_2(\vec{k})$, $\vec{a_1} \rightarrow -\vec{a_2}$, $\vec{a_2} \rightarrow -\vec{a_1}$, and the above relation becomes $(E^4-5E^2+3) \psi_2(\vec{k}) =  (e^{i\vec{k} \cdot \vec{a_1}} + e^{i\vec{k} \cdot \vec{a_2}} + 1) \psi_5(\vec{k})$. For the two relations to hold simultaneously, we require that
\begin{equation}
(E^4-5E^2+3)^2 \psi_5(\vec{k}) =  |e^{i\vec{k} \cdot \vec{a_1}} + e^{i\vec{k} \cdot \vec{a_2}} + 1|^2 \psi_5(\vec{k}).
\label{eq:check}
\end{equation}

On the left hand side of Eq. \eqref{eq:check}, when $\psi(\vec{k})$ is a flat band wave function, its energy $E$ is independent of $\vec{k}$ over the entire BZ. However, on the right hand side, the coefficient in front of $\psi_5(\vec{k})$ is $\vec{k}$-dependent. Therefore, for Eq. \ref{eq:check} to be satisfied for all $\vec{k}$, we have to have $\psi_5(\vec{k})=0$ over the BZ. Furthermore, since $\psi_2(\vec{k}) \propto \psi_5(\vec{k})$, $\psi_2(\vec{k}) = 0$ as well.  
In addition, if $\psi_5 \neq 0$, solving the eighth-order polynomial equation $(E^4-5E^2+3)^2 = |e^{i\vec{k} \cdot \vec{a_1}} + e^{i\vec{k} \cdot \vec{a_2}} + 1|^2$ from Eq. \eqref{eq:check} at $\vec{k} = \vec{0}$, we would obtain the eight \textit{dispersive} band energies, $E = 0,0,\pm \sqrt{2},\pm \sqrt{3}, \pm \sqrt{5}$. This is consistent with our numeric band calculation at the $\Gamma$ point in Fig. \ref{fig:bandstructures}(a).
Similar results can be obtained for the honeycomb superlattice with a different number of additional sites along the honeycomb edge. 


\section{\label{sec:fb}Flat Bands with  $C_{3v}$-symmetric hopping perturbations}
To further study the relation between flat bands and lattice symmetry, 
we introduce additional, longer range hoppings in the honeycomb superlattice model that preserve the lattice $C_{3v}$ symmetry. 
Remarkably, the three flat bands found in the previous section all survive in the presence of these symmetric perturbations [\onlinecite{GYC}]. 
Below, we explain our analysis of this result, analytically derive the three flat band energies, and further discuss the tunability of these flat bands.

As shown in Fig. \ref{fig:model}(b), we introduce additional 
long-range hoppings with amplitudes $t_1$, $t_2$ and $t_3$ 
that are all $C_3$-symmetric about the honeycomb vertices.  
When all these three types of hoppings are present,
the Hamiltonian matrix kernel takes the form
\begin{equation}
{\mathcal{H}}(\vec{k}) =  
\begin{bmatrix}
{\mathcal{H}}_{AA}(t_2, \vec{k})  & {\mathcal{H}}_{BA}(t_0, t_3, \vec{k}) \\
{\mathcal{H}}_{AB}(t_0, t_3, \vec{k})  & {\mathcal{H}}_{BB}(t_1, \vec{k}) \\
\end{bmatrix}.
\label{eq:t1matrix}
\end{equation}
Here, $t_{1}$ and $t_2$ contribute to the diagonal blocks ${\mathcal{H}}_{BB}$ and ${\mathcal{H}}_{AA}$, respectively, in Eq. \eqref{eq:t1matrix}. 
Since $t_3$ involves hopping between $A$ and $B$ sublattice sites, it modifies the off-diagonal blocks ${\mathcal{H}}_{AB}$ and  ${\mathcal{H}}_{BA}$.  
The explicit expressions of blocks ${\mathcal{H}}_{ij}$ with $i, j=A, B$ in Eq. \eqref{eq:t1matrix} are given by 
\begin{equation}
{\mathcal{H}}_{AA} = -t_2
\begin{bmatrix}
  0 & 0 & 1+ \omega_1^{*} \omega_2  & 1+\omega_1^{*} & 0  \\
  0 & 0 & 0 & 0 & 0 \\
  1+\omega_1 \omega_2^{*}  & 0 & 0 & 1+\omega_2^{*} & 0 \\
  1+\omega_1 & 0 & 1 + \omega_2 & 0 & 0 \\
  0 & 0 & 0 & 0 & 0 \\
  \end{bmatrix}\\,
\end{equation}
\begin{equation}
    {\mathcal{H}}_{BB} = -t_1 \begin{bmatrix}
  0 & 1 & 1 & 0 & 0 & 0 \\
  1 & 0 & 1 & 0 & 0 & 0 \\
  1 & 1 & 0 & 0 & 0 & 0 \\
  0 & 0 & 0 & 0 & 1 & 1\\
  0 & 0 & 0 & 1 & 0 & 1\\
  0 & 0 & 0 & 1 & 1 & 0
  \end{bmatrix}\\,
\end{equation}
and
\begin{equation}
    {\mathcal{H}}_{AB} = \begin{bmatrix}
  t_0 & t_0 & t_3 & t_3 & 0  \\
  t_3 & t_0 & t_0 & t_3 & 0  \\
  t_3 & t_0 & t_3 & t_0 & 0 \\
  t_3 \omega_1 & 0 & t_3 \omega_2 & t_0 & t_0\\
  t_3 \omega_1 & 0 &  t_0 \omega_2 & t_3 & t_0\\
  t_0 \omega_1 & 0 & t_3 \omega_2 & t_3 & t_0
  \end{bmatrix}
  = \mathcal{H}_{BA}^\dagger\\.
\end{equation}
Here, $\omega_1(\vec{k})=e^{i \vec{k} \cdot \vec{a_1}}$ and  
$\omega_2(\vec{k})=e^{i \vec{k} \cdot \vec{a_2}}$, and since $t_1$ and $t_2$ are nonbipartite hoppings, $H_{AA}$ and $H_{BB}$ are written with $t_1, t_2 > 0$, which cannot be changed by a gauge transformation.

We can first derive the relation between $\psi_2(\vec{k})$ and $\psi_5(\vec{k})$ as
$[E^4+(2t_1+t_2)E^3-(5+2t_3)E^2+(-2t_1+3t_2-2t_1t_2)E+3] \psi_2(\vec{k}) 
=(4t_3+1) (1+\omega_1(\vec{k})+\omega_2(\vec{k})) \psi_5(\vec{k})$.
Using flat bands' momentum independence and the system's inversion symmetry, as in Sec. \ref{sec:model}, we can derive that the flat band Bloch wave function also vanishes at the honeycomb vertices in the presence of these longer range hoppings. Therefore, we next use the ansatz $\psi(\vec{k}) = [ \psi_1(\vec{k}), 0, \psi_3(\vec{k}), \psi_4(\vec{k}), 0, \psi_6(\vec{k}) \cdots, \psi_{11}(\vec{k})] ^{T}$
to solve the flat band energies from the eigenequations of ${\mathcal{H}}(\vec{k})$ in Eq. \eqref{eq:t1matrix}. 
These equations can be reduced to
\begin{eqnarray}
(E-t_1)(E-2t_2) \psi_i(\vec{k}) 
&=& 2 (1-t_3)^2 \psi_i(\vec{k}), \label{eq:psi134} \\
\psi_1(\vec{k}) + \psi_3(\vec{k}) + \psi_4(\vec{k}) &=& 0. \label{eq:constraint}
\end{eqnarray}
with $ i=1,3,4$ in Eq. \ref{eq:psi134} and the remaining components being linear combinations of $\psi_1(\vec{k}),$ $\psi_3(\vec{k}),$ and $\psi_4(\vec{k})$, as shown in Appendix \ref{app:H_eigenequation}.
Thus, Eq. \eqref{eq:constraint} requires that
\textit{at most one} of $\psi_1(\vec{k})$, $\psi_3(\vec{k})$, and  $\psi_4(\vec{k})$ is allowed to vanish at any $\vec{k}$. Therefore, at least two of the three equations in Eq. \eqref{eq:psi134} must hold non-trivially and we must have 
\begin{equation}
    (E-t_1) (E-2t_2)= 2 (t_3-1)^2.
\label{eq:fbenergy}
\end{equation}
In terms of $\Delta \equiv (t_1/2-t_2)^2 + 2(t_3-1)^2\ge 0$, the two roots of Eq. \eqref{eq:fbenergy} are $(t_1/2+t_2)\pm \sqrt{\Delta}$ which correspond to $E_{\mathrm{flat,top/bot}}$. The middle flat band occurs at $E = t_1$. 

Eq.  \eqref{eq:fbenergy} 
is consistent with the flat band energies we obtained when only the N.N. hoppings are present, where $\Delta=2$,  $E_{\mathrm{flat, top/bot}} = \pm \sqrt{2}$ and  $E_{\mathrm{flat, mid}} =0$, in agreement with Fig. \ref{fig:bandstructures}(a).  
When only $t_1$ is nonzero with $t_1 = 0.3$, 
Eq. \eqref{eq:fbenergy} gives flat band energies $E_{\mathrm{flat, top/bot}} = 0.15 \pm \sqrt{2+(0.15)^2}$ which take approximate values $1.57$ and  $-1.27$, 
and  $E_{\mathrm{flat, mid}} = t_1 = 0.3$ as shown in Fig. \ref{fig:bandstructures} (b). 
When only $t_2$ is nonzero with  $t_2 = 0.3$, $E_{\mathrm{flat, top/bot}} = 0.3 \pm \sqrt{2+(0.3)^2}$ which take approximate values $1.75$ and $-1.15$, and  $E_{\mathrm{flat, mid}} = 0$ as shown in Fig. \ref{fig:bandstructures} (c).
Finally, when only $t_3$ is nonzero with  $t_3 = 0.3$, $E_{\mathrm{flat, top/bot}} = \pm \sqrt{2 \times (0.7)^2}$ which take approximate values $0.99$ and $-0.99$, and  $E_{\mathrm{flat, mid}} = 0$ as shown in Fig. \ref{fig:bandstructures} (d). This is consistent with the fact that $t_3$ preserves particle-hole symmetry of the system.

With the explicit formulas, the flat band energies are \textit{highly tunable} as the values of the further-neighbor hopping amplitudes vary. For example, if we wish to obtain three flat bands at given values of $E_1$, $E_2$, and $E_3$ with $E_1 < E_2 < E_3$, 
we can first obtain the middle flat band by tuning $t_1 = E_2$. Then, we solve for $t_2$ and $t_3$ that will make $E_1$ and $E_3$ the roots to Eq. \eqref{eq:fbenergy}. Using Vieta's formula, we need
\begin{equation}
\begin{gathered}
t_2 = \frac{E_1+E_3-t_1}{2} \\
(1-t_3)^2 = \frac{(- E_1+t_1 )(E_3 - t_1)}{2}\\
\end{gathered}
\end{equation}
Note that since $E_2 = t_1$, $-E_1+t_1 > 0$, and $E_3 - t_1 > 0$, the RHS is always positive, and one can always find an appropriate value of $t_3$ to realize any desired values of flat band energies at $E_1$ and $E_3$.


\section{\label{sec:localizedstates}Eigenfunctions of flat bands in Real Space}
In this section, we analyze the structure of wave functions in real space. We first show that the wave functions on each honeycomb edge are standing waves at different energies, which explains the existence of multiple flat bands. Furthermore, we construct localized states on the honeycomb plaquettes and argue that they remain localized in the presence of longer range hoppings, thereby accounting for the interesting robustness of flat bands.

We start with the system with only N.N. hoppings described by $H_{\mathrm{N.N.}}$ in Eq. \eqref{eq:Hamiltonian}. 
From our analysis in Sec. \ref{sec:model}, the wave functions of the flat bands vanish at the honeycomb vertices. In real space, this occurs due to destructive interference when the wave function has alternating signs on neighboring honeycomb edges. 
We consider first the structure of the wave function on a honeycomb edge with five sites assuming the wave function vanishes at the honeycomb vertices in the flat bands. Sequentially labeling the wave function components along the edge $\phi_0, \phi_1, \phi_2, \phi_3, \phi_4$, the components on the ends vanish, $\phi_0 = \phi_4 = 0$, and the remaining components satisfy
\begin{eqnarray}
    E \phi_1 &=& \phi_2, \nonumber \\
    E \phi_2 &=& \phi_1 + \phi_3, \nonumber \\
    E \phi_3 &=& \phi_2.
\label{eq:realeigeq}
\end{eqnarray}
The solutions of Eq. \eqref{eq:realeigeq} describe standing waves with $\phi_{n}^{(m)} = C \sin\left(\frac{m \pi}{4} n \right)$, where $m \in \{1,2,3\}$ labels the band, $n \in \{0,1,2,3,4\}$ labels the site, and $C$ is an overall normalization constant. The corresponding energies are
\begin{equation}
    E_m = 2 \cos\left(\frac{m \pi}{4}\right).
    \label{eq:realenergy}
\end{equation}
Eq. \eqref{eq:realenergy} agrees with the three flat band energies at $E = \pm \sqrt{2}, 0$ obtained in Sec. \ref{sec:fb} and in Fig. \ref{fig:bandstructures}(a). 

From the wave function's standing wave structure, we can understand why the system in general can host multiple flat bands. 
Following the same reasoning in Sec. \ref{sec:model}, we see the wave functions vanish at the junctions due to symmetry. 
When there are $N$ sites on each honeycomb edge, 
each edge can therefore host $N-2$ possible standing wave solutions.  
In particular, $\phi_{n}^{(m)} \sim \sin ( \frac{m \pi}{N-1} n)$, where $n \in \{0,1, \cdots, N-1 \}$ labels the site, and $m$ labels the band. 
We see that the allowed  values of $m$ run from 1 to  $N-2$, each corresponding to one flat band. 
This accounts for the observation that the number of flat bands is proportional to the number of sites along the edge [\onlinecite{GYC}].

\begin{figure}[tpb]
\subfigure[]{\includegraphics[width=0.6\linewidth]{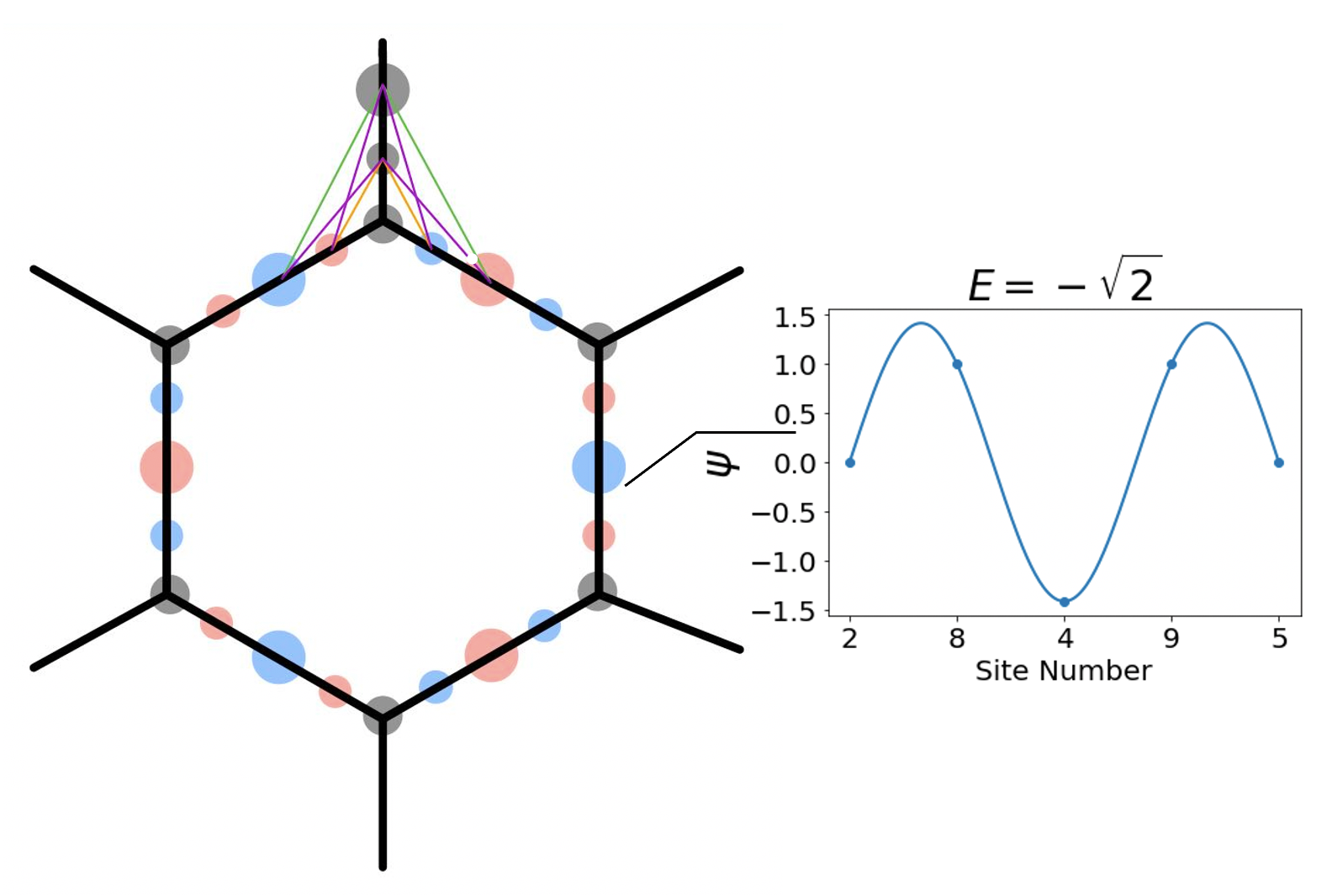}}
\subfigure[]{\includegraphics[width=0.6\linewidth]{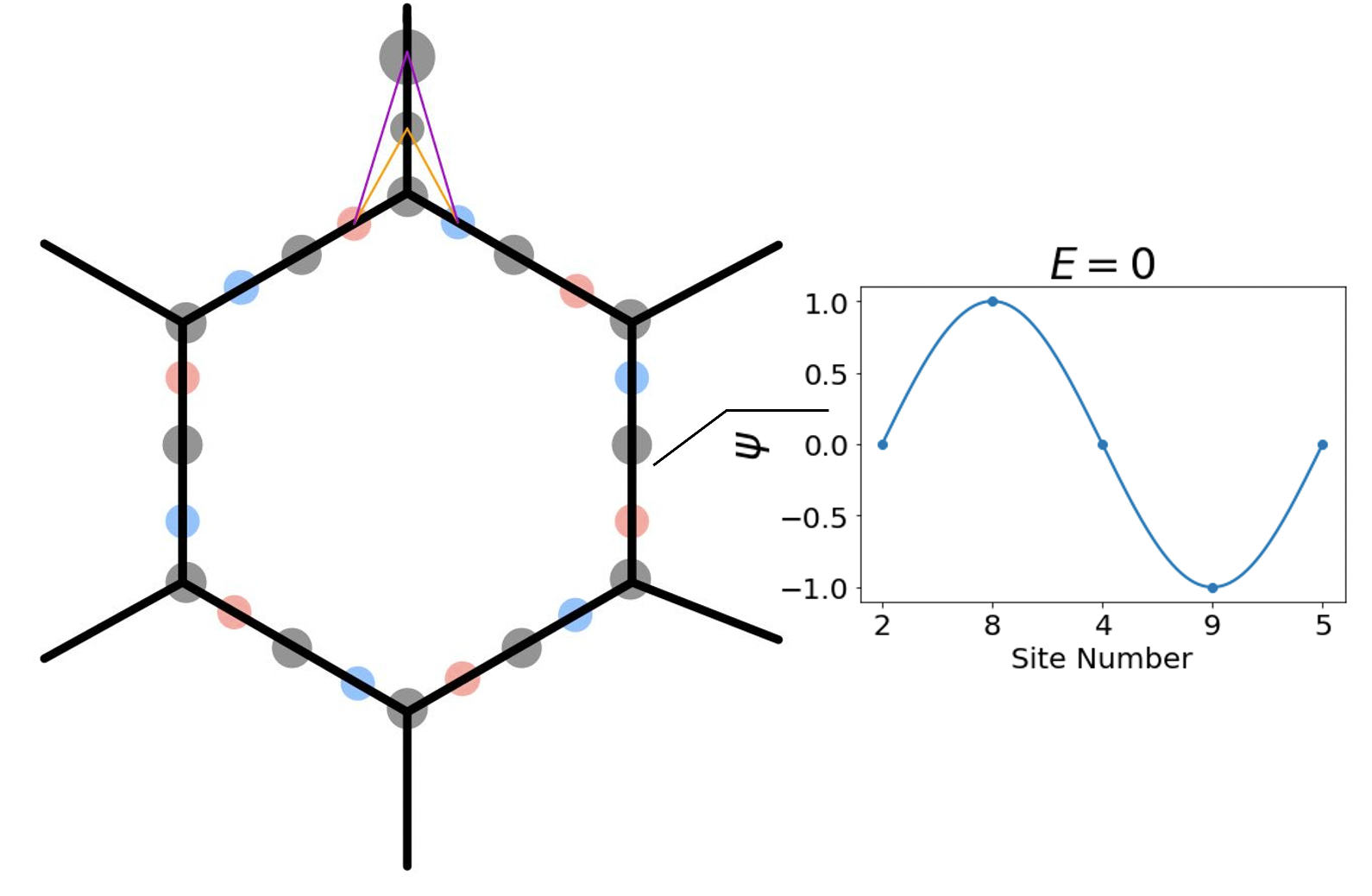}}
\subfigure[]{\includegraphics[width=0.6\linewidth]{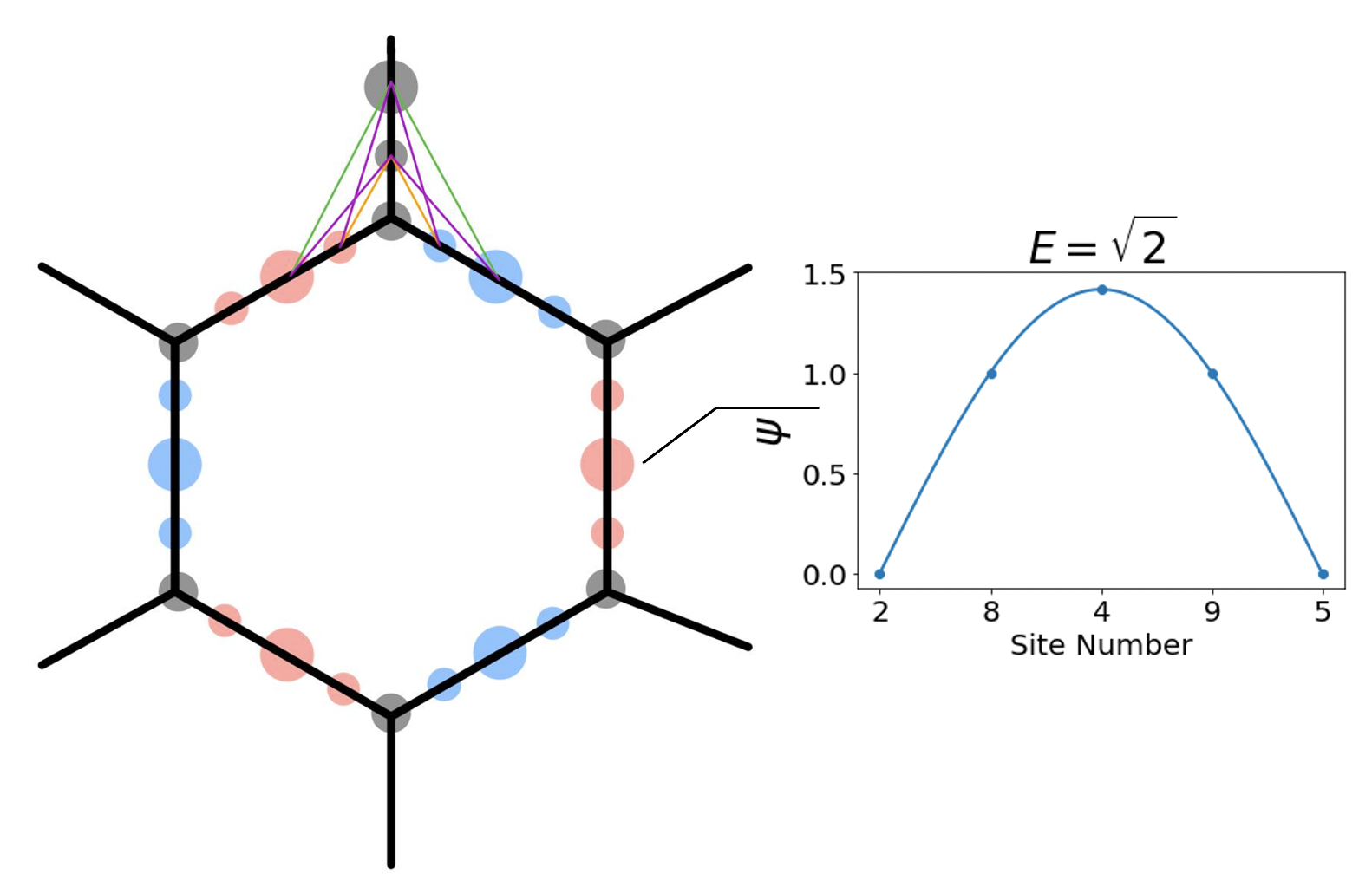}}
\caption{Localized eigenstates around a plaquette for the flat band at (a)$E_{\mathrm{flat, bot}}  = -\sqrt{2}$, (b)$E_{\mathrm{flat, mid}} =0$, 
(c) $E_{\mathrm{flat, top}} =\sqrt{2}$. The wave functions on sites in red, blue, and black have positive-, negative-, and zero-valued weights, respectively (up to an overall sign). 
As shown around the top honeycomb vertices, for each type of perturbation, the hopping from any two neighboring edges always cancel out due to destructive interference in real space, rendering the plaquette states Wannier-like localized. The insets show the structures of the wave functions living on the vertical edge containing sites 2,8,4,9,5, which are well fit with standing waves.}
\label{fig:localizedstates}
\end{figure}

\begin{figure}[tpb]
{\includegraphics[scale=0.2]{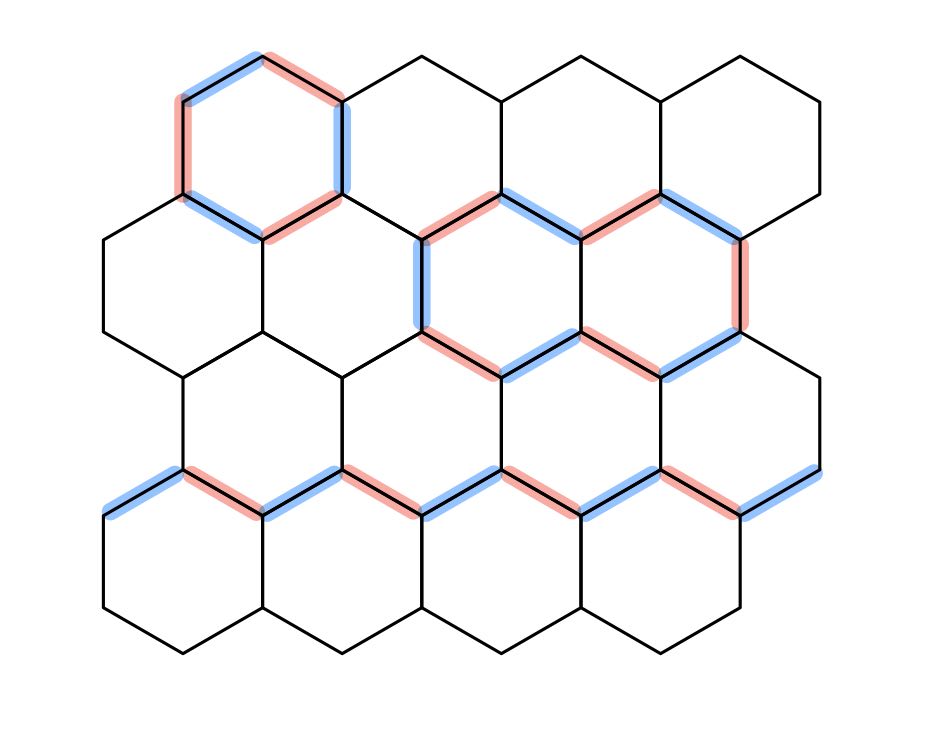}}
\caption{States localized on hexagonal plaquettes and loops in the honeycomb superlattice. Wave function on bonds labelled in red and blue have opposite signs. 
Each plaquette has the same configuration as one of the three shown in Fig. \ref{fig:localizedstates}, depending on the flat band energy. Larger loop states formed by superposing multiple plaquette states are contractible. When periodic boundary conditions are imposed, non-contractible loop states can be constructed winding around the entire lattice at the same energy as other flat-band eigenstates.
}
\label{fig:wannier}
\end{figure}

Flat band wave functions in real space can be found by piecing together same-energy standing waves along each honeycomb edge.
In particular, plaquette states \cite{GYC} can be formed from standing waves taking alternating signs around the honeycomb edges of a plaquette,
\begin{equation}
|\psi^{(m)}_{\varhexagon}\rangle = \sum_{\delta=1}^6\sum_{n = 0}^4 (-1)^{\delta} \phi_{n}^{(m)} c^\dagger_{\vec{r}_{\delta,n}}|0\rangle
\end{equation}
where $\delta$ labels the edges of the hexagonal plaquette counterclockwise from the right and $n_{\delta}$ labels the $n$th site on edge $\delta$, with $\vec{r}_{\delta, n}$ the corresponding real space vector. 
The localized plaquette states for each of the three flat bands when $N=5$ are shown in Fig. \ref{fig:localizedstates}.
Similar plaquette states are eigenstates of the flat bands
in the presence of longer range hoppings respecting lattice symmetries. 
Due to destructive interference from standing waves of opposite signs on any two neighboring edges, such states are always localized within plaquettes without ``leaking'' off to any sites outside a plaquette, as shown schematically in Fig. \ref{fig:wannier}  as Wannier-like localized states. By superposing neighboring honeycomb plaquettes, one can form larger contractible loop states over the lattice. In the presence of periodic boundary conditions, one can construct non-contractible loop states that wind around the entire lattice and are flat band eigenstates.
All these states are localized on either plaquettes or loops in the lattice and therefore are macroscopically degenerate, thereby giving rise to the flat bands. Since the localized states are not dispersed by additional longer range hoppings, the flat bands remain robust and tunable in the presence of symmetric perturbations. 

\section{\label{sec:conclusion}Conclusions}
In this paper, we studied the flat bands arising from honeycomb superlattices without and with longer range hoppings respecting the lattice symmetry. We showed that the presence of flat bands is robust with and the flat band energies are tunable in terms of different hopping strengths.
We also analyzed the wave function structure and explicitly constructed eigenstates of these flat bands that remain localized due to destructive interference. 
As long as the longer range hoppings preserve the destructive interference based on the sign structure of the eigenfunctions, 
these states remain localized around plaquettes and loops in the lattices, accounting for the flat bands with macroscopic degeneracy and tunable energies.

\section{Acknowledgements}
We thank G.-Y. Cho for bringing the works by him and his collaborators to our attention. This work is supported by the NSF CAREER grant DMR-1848349 and in part by Alfred P. Sloan Research Fellowships under grant FG-2018-10971. 
Z.Q. is also grateful for Caltech's Summer Undergraduate Research Fellowship (SURF).

\appendix
\section{Flat band eigenequations with longer range hoppings $t_1$, $t_2$ and $t_3$}
\label{app:H_eigenequation}
We consider the equations for the flat band states of ${\mathcal{H}}(\vec{k})$ in Eq. \eqref{eq:t1matrix}. Noting that flat band states have vanishing components $\psi_2(\k)$ and $\psi_5(\k)$, the flat band states can be written $\psi(\vec{k}) = [ \psi_1(\vec{k}), 0, \psi_3(\vec{k}), \psi_4(\vec{k}), 0, \psi_6(\vec{k}) \cdots, \psi_{11}(\vec{k})] ^{T}$. For convenience, the momentum dependence of $\psi_i \equiv \psi_i(\vec{k})$ will be suppressed in what follows.

The eigenvalue equations for $\psi_2$ and $\psi_5$ yield the relations
\begin{equation}
\begin{gathered}
\psi_6 + \psi_7 + \psi_8 = 0, \\
\psi_9 + \psi_{10} +\psi_{11} = 0.
\end{gathered}
\label{appeq:important}
\end{equation}
Using these relations, the eigenvalue equations for the remaining components become 
\begin{eqnarray}
E\psi_1 
&=& -t_2[(1+e^{-i \vec{k} \cdot \vec{(a_1-a_2)}})\psi_3 + (1+e^{-i \vec{k} \cdot \vec{a_1}}) \psi_4 ] \nonumber \\ 
&&+(1-t_3)(\psi_6 + e^{-i\vec{k} \cdot \vec{a_1}} \psi_{11}), \label{appeq:eigen_1}\\
E\psi_3 
&=&  -t_2[(1+e^{i \vec{k} \cdot \vec{(a_1-a_2)}})\psi_1 + (1+e^{-i \vec{k} \cdot \vec{a_2}}) \psi_4]\nonumber \\ 
&&+(1-t_3)(\psi_7 + e^{-i\vec{k} \cdot \vec{a_2}} \psi_{10}),\label{appeq:eigen_3}\\
E\psi_4 
&=&  -t_2[(1+e^{i \vec{k} \cdot \vec{a_1}})\psi_1 + (1+e^{i \vec{k} \cdot \vec{a_2}}) \psi_3] \nonumber \\ 
&&+(1-t_3)(\psi_8 + \psi_{9}), \label{appeq:eigen_4}\\
(E-t_1) &\psi_6& = (1-t_3) \psi_1, \label{appeq:eigen_6}\\
(E-t_1) &\psi_7& = (1-t_3) \psi_3, \label{appeq:eigen_7}\\
(E-t_1) &\psi_8& = (1-t_3) \psi_4,\label{appeq:eigen_8} \\
(E-t_1) &\psi_9& = t_3 e^{i \vec{k} \cdot \vec{a_1}} \psi_1 + t_3 e^{i \vec{k} \cdot \vec{a_2}} \psi_3 + \psi_4, \label{appeq:eigen_9}\\
(E-t_1) &\psi_{10}& = t_3 e^{i \vec{k} \cdot \vec{a_1}} \psi_1 + e^{i \vec{k} \cdot \vec{a_2}} \psi_3 + t_3 \psi_4, \label{appeq:eigen_10}\\
(E-t_1) &\psi_{11}& = e^{i \vec{k} \cdot \vec{a_1}} \psi_1 + t_3 e^{i \vec{k} \cdot \vec{a_2}} \psi_3 + t_3 \psi_4. \label{appeq:eigen_11}
\label{eq:modifiedrelations}
\end{eqnarray}
 
We note immediately that $E = t_1$ is a solution for a non-zero eigenstate. 
Next, we derive the other two flat band energies. By adding Eqs. \eqref{appeq:eigen_6}, \eqref{appeq:eigen_7} and \eqref{appeq:eigen_8} and using Eq. \eqref{appeq:important}, we have $\psi_1 + \psi_3 + \psi_4 = 0$. 
Following the same procedure, from Eqs. \eqref{appeq:eigen_9}, \eqref{appeq:eigen_10} and \eqref{appeq:eigen_11} with $E \neq t_1$, we have 
\begin{equation}
\psi_4 + e^{i\vec{k} \cdot \vec{a_2}} \psi_3 + e^{i\vec{k} \cdot \vec{a_1}} \psi_1 = 0.
\label{eq:simplify}
\end{equation}
Finally, applying Eqs. \eqref{appeq:eigen_9} and \eqref{appeq:eigen_10} to the right hand side of Eq. \eqref{appeq:eigen_1}  and using Eq. \eqref{eq:simplify} to simplify, we get 
\begin{equation}
E \psi_1 = -2-t_2 \psi_1 + \frac{(1-t_3)(2-2t_3)}{E-t_1} \psi_1.
\label{eq:psi1}
\end{equation}
Following the same procedure, we obtain two similar relations for $\psi_3$ and $\psi_4$,
\begin{equation}
\begin{gathered}
E \psi_3 = -2-t_2 \psi_3 + \frac{(1-t_3)(2-2t_3)}{E-t_1} \psi_3,  \\
E \psi_4 = -2-t_2 \psi_4 + \frac{(1-t_3)(2-2t_3)}{E-t_1} \psi_4.
\label{eq:psi34}
\end{gathered}
\end{equation}


%

\end{document}